\documentclass[conference]{IEEEtran}
\usepackage{fontspec}
\usepackage{xeCJK}
\usepackage{amsmath,amssymb}
\usepackage{graphicx}
\usepackage{booktabs}
\usepackage{multirow}
\usepackage{xcolor}
\usepackage{url}
\usepackage{tikz}
\usetikzlibrary{positioning,arrows.meta}

\begin{document}

\title{Reading-Level Verification of Kanji Words\\in Japanese Financial Speech Synthesis}

\author{\IEEEauthorblockN{Cheng ZHANG}
\IEEEauthorblockA{Best AI Co., Ltd., Tokyo, Japan\\
Email: bestaijapan@gmail.com}}

\maketitle

\begin{abstract}
Few-shot voice cloning makes it cheap to produce daily Japanese financial commentary with several synthetic speakers, but kanji words with more than one reading, such as 市場 (shij\=o vs.\ ichiba) or 大勢 (taisei vs.\ \=oze), are still misread. A common quality gate transcribes each synthesized sentence with automatic speech recognition (ASR), converts both the input and the transcript to kana, and regenerates on mismatch. We show that this gate is unreliable in both directions: when the audio contains a misreading, the ASR language model frequently writes the expected kanji back, hiding the error, and when the audio is correct, ordinary recognition errors trigger needless regeneration. We propose a reading-level check that scores the expected reading and its confusable alternatives with a connectionist temporal classification (CTC) hiragana recognizer and flags a word when an alternative explains the audio better. On 540 utterances from three voice-cloned production voices covering 30 financial terms, with misreadings injected under controlled conditions, the transcript-based gate detected 78.9\% of misreadings at a 44.4\% false-alarm rate, whereas the proposed check detected 91.7\% at 5.0\% (AUC 0.990). On natural kanji input, it caught a front-end error that the transcript-based gate missed for one voice, and reduced regeneration requests from 38.9\% to 6.1\% of sentences.
\end{abstract}


\section{Introduction}
Synthetic voices are increasingly used to produce daily market commentary for video platforms and podcasts, often as a dialogue between two characters. Few-shot voice cloning systems built on neural codec or semantic-token language models~\cite{valle,speartts} allow a small team to maintain several distinct voices and to generate a ten-minute episode in a few minutes. In this setting the main quality problem is no longer naturalness but \emph{correctness}. Listeners of financial content are sensitive to terminology: a voice that reads 市場 (market) as \emph{ichiba} (marketplace) or 大勢 (overall trend) as \emph{\=oze} (a crowd) sounds amateurish and damages the credibility of the channel.

Japanese is particularly prone to such errors because many kanji words have several readings whose choice depends on domain and context. Financial text adds a vocabulary of jargon with idiosyncratic readings (上値 \emph{uwane}, 後場 \emph{goba}, 手仕舞い \emph{tejimai}) and of numeral expressions whose euphonic changes are easy to get wrong (十分 \emph{juppun} ``ten minutes'' vs.\ \emph{j\=ubun} ``enough''). Misreadings come from two sources: the text front-end may predict the wrong reading, and the acoustic model, fine-tuned on limited speaker data, may deviate from the phonemes it was given. In our production system we have observed both, e.g., 千円 rendered as \emph{sennen} in all of nine candidate generations although the front-end reading was correct.

A practical and widely used safeguard is an ASR-based gate: synthesize each sentence, transcribe it with a robust recognizer such as Whisper~\cite{whisper}, compare the transcript with the input, and regenerate on mismatch. Comparing kana readings rather than surface strings avoids false alarms from orthographic variants (上げ/揚げ). Our own pipeline uses exactly this design. However, the gate relies on the recognizer's \emph{text} output, and a strong recognizer is also a strong language model. When the audio says \emph{ichiba} in a sentence about stocks, the most probable transcript is still 市場, which the morphological analyzer converts back to the expected \emph{shij\=o}; the gate passes and the error reaches the listener. Conversely, when the audio is correct, recognition errors on rare jargon change the kana string and cause needless regeneration.

This paper makes three contributions:
\begin{itemize}
\item We quantify, under controlled error injection, how often the transcript-based kana gate misses misreadings and how often it raises false alarms on correct audio, using three production voices and 30 financial terms.
\item We propose a reading-level check that compares CTC likelihoods of the expected reading and its confusable alternatives, independent of the recognizer's choice of kanji.
\item We show that the check also detects a real front-end misreading in natural generation and sharply reduces unnecessary regeneration.
\end{itemize}

\section{Related Work}
\textbf{Reading prediction.} Japanese TTS front-ends predict readings with a morphological analyzer and dictionary, as in Open JTalk~\cite{openjtalk} and its Python binding pyopenjtalk~\cite{pyopenjtalk}, optionally with user dictionaries. Neural approaches to heteronym and polyphone disambiguation, e.g., g2pW for Mandarin~\cite{g2pw}, improve accuracy on general text. All of these act \emph{before} synthesis and cannot observe errors introduced by the acoustic model, nor verify that a domain term was spoken as intended.

\textbf{Neural TTS and voice cloning.} End-to-end models such as VITS~\cite{vits} synthesize speech directly from phonemes, while recent few-shot systems treat TTS as language modeling: VALL-E~\cite{valle} predicts neural codec tokens and SPEAR-TTS~\cite{speartts} predicts self-supervised semantic tokens autoregressively from text, before a second stage generates the waveform. Such two-stage systems build custom voices from small amounts of speech, but their autoregressive stage can deviate from the input phonemes, and for Japanese they still depend on a front-end for readings.

\textbf{ASR for TTS evaluation.} ASR transcripts are routinely used to measure the intelligibility of synthesized speech via character or word error rate, and large models such as Whisper~\cite{whisper} and WhisperX~\cite{whisperx} are popular for this purpose. Such metrics measure what the recognizer \emph{writes}, which for logographic scripts is only loosely tied to what was \emph{pronounced}.

\textbf{Pronunciation assessment.} Scoring the likelihood of an expected phone sequence against alternatives is the basis of goodness-of-pronunciation measures in computer-assisted language learning~\cite{gop}. We apply the same idea to synthesized speech, using a CTC~\cite{ctc} recognizer with kana outputs and a small set of confusable readings per word.

\section{System}

\begin{figure}[t]
\centering
\resizebox{\columnwidth}{!}{%
\begin{tikzpicture}[node distance=2.2mm, every node/.style={font=\scriptsize}, box/.style={draw, rounded corners=1pt, align=center, minimum height=5.5mm, inner sep=2pt}, >={Stealth[length=1.6mm]}]
\node[box] (txt) {Script\\(speaker, style, text)};
\node[box, right=of txt] (sub) {Substitution\\table};
\node[box, right=of sub] (tts) {TTS per\\sentence};
\node[box, below=5mm of tts] (s1) {Stage 1: Whisper\\+ kana compare};
\node[box, left=of s1] (s2) {Stage 2: CTC scoring\\of candidate readings};
\node[box, left=of s2] (cat) {Concatenate\\(pauses)};
\draw[->] (txt) -- (sub); \draw[->] (sub) -- (tts); \draw[->] (tts) -- (s1);
\draw[->] (s1) -- node[above]{pass} (s2); \draw[->] (s2) -- node[above]{pass} (cat);
\draw[->] (s1.east) -- ++(3mm,0) |- node[pos=0.25, right]{fail} (tts.east);
\draw[->] (s2.south) -- ++(0,-2.5mm) -| node[pos=0.75, right]{fail: retry $\le K$} ([xshift=6mm]tts.east);
\end{tikzpicture}}
\caption{Generate--verify--regenerate loop. Stage~1 is the deployed transcript-based gate; Stage~2 is the proposed reading-level check.}
\label{fig:pipeline}
\end{figure}
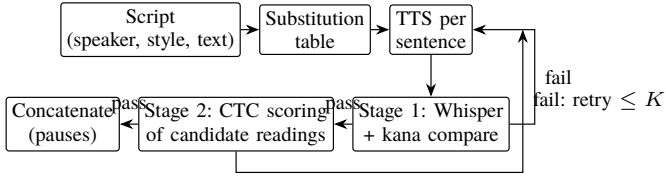

\subsection{Generation Pipeline}
Fig.~\ref{fig:pipeline} shows the production loop. Scripts are written as dialogue; each line carries a speaker and a speaking-style tag. Lines of the same speaker are synthesized together to avoid repeated model loading and are concatenated in script order with 0.3\,s pauses between sentences and 0.6\,s at speaker or paragraph changes. The voices are built on an open-source two-stage voice-cloning framework of the type described in Section~II: pyopenjtalk converts the text to phonemes, an autoregressive Transformer predicts self-supervised semantic tokens from the phonemes, and a VITS-based decoder~\cite{vits} generates the waveform. Each voice is fine-tuned with LoRA~\cite{lora} on approximately 30 minutes of Japanese speech from a single speaker and conditioned on a 3--10\,s reference clip per style.

Before synthesis, a substitution table rewrites expressions that have been misread in the past into kana (e.g., 円安 $\rightarrow$ えんやす, 千円 $\rightarrow$ せんえん). Entries are applied longest-match first, may be restricted to particular voices, and numeral-initial entries are skipped when preceded by a digit. The original text is kept for subtitles and verification. The table is reactive by design: an entry is added only after an error has been heard, which is why automatic detection matters.

\subsection{Stage 1: Transcript-Based Kana Check}
Each synthesized sentence is transcribed by WhisperX (large-v3)~\cite{whisperx}. Both the input text and the transcript are converted to katakana readings with pyopenjtalk, all non-kana symbols are removed, and the sentence passes only if the two strings are identical. Orthographic variants therefore count as matches. This stage reliably catches skipped, repeated or hallucinated phrases, which are typical failure modes of autoregressive TTS, and misreadings that lead the recognizer to a different word.

\subsection{Stage 2: Reading-Level Check}
Stage~2 targets words with known confusable readings. For each such word $w$ in a sentence, let $r^*$ be the expected reading, taken from a domain lexicon, and $\mathcal{A}(w)$ a small set of alternative readings that a front-end or acoustic model could plausibly produce (in this study, one alternative per word; Table~\ref{tab:words}). The sentence is split as $x = p\,w\,q$, and the hiragana readings $\rho(p)$ and $\rho(q)$ of the surrounding text are taken from the front-end. For a candidate reading $r$, the sentence-level label sequence is $y(r) = \rho(p)\,r\,\rho(q)$, and its score is the CTC log-likelihood
\begin{equation}
s(r) = \log P_{\mathrm{CTC}}\bigl(y(r) \mid \mathbf{a}\bigr),
\end{equation}
where $\mathbf{a}$ is the audio and $P_{\mathrm{CTC}}$ is computed by a wav2vec~2.0 XLSR-53 model~\cite{xlsr} fine-tuned on Common Voice Japanese~\cite{commonvoice} with hiragana targets~\cite{w2v2hira}. The word is flagged if
\begin{equation}
m(w) = \max_{r \in \mathcal{A}(w)} s(r) - s(r^*) > \theta ,
\label{eq:margin}
\end{equation}
and the sentence fails if any word is flagged. Because $y(r)$ and $y(r^*)$ differ only in the target span, context errors of the recognizer affect both scores similarly and largely cancel in the margin $m(w)$. The decision depends on acoustic evidence for each reading rather than on which kanji a language model would write. The cost is one forward pass of the CTC model per sentence plus one CTC loss evaluation per candidate.

\subsection{Building the Reading Lexicon}
Stage~2 needs, for each word to be verified, its expected reading and a few confusable alternatives. We build this lexicon from three sources. First, every entry added to the substitution table records an observed misreading, which becomes an alternative for that word. Second, domain glossaries of market terms provide the expected financial readings, which often differ from the general-domain default (e.g., 上値 \emph{uwane}, 後場 \emph{goba}). Third, alternatives can be enumerated from dictionary readings of the constituent kanji and from the front-end's own output when it disagrees with the glossary; this third step is not used in the present study, where one alternative per word was chosen manually. Because the check only scores readings and does not need audio examples, a new term can be covered by adding one line to the lexicon.

The expected reading must not be taken from the same front-end that drives the synthesizer: if the front-end is wrong, synthesizer and verifier agree on the wrong reading. Section~\ref{sec:natural} shows a real instance of this failure.

\subsection{Regeneration Policy}
A sentence failing either stage is regenerated with a new random seed up to $K$ times ($K\!=\!9$ in production: three rounds of three candidates), and the first candidate passing both stages is kept. If none passes, the best-scoring candidate is used and the flagged word is reported, so that the operator can add an entry to the substitution table.

\section{Experimental Setup}

\subsection{Voices and Test Set}
We used three production voices: two female (F1, F2) and one male (M1), fine-tuned from the framework's pretrained models with LoRA ranks 64, 32 and 32, respectively. Each voice was conditioned on its default reference clip.

We selected 30 terms frequent in Japanese market commentary, each with one financial reading and one plausible misreading (Table~\ref{tab:words}). The misreadings fall into three types: (a) \emph{reading choice}, where a different on- or kun-reading of the same characters is used (21 terms, e.g., 市場 \emph{ichiba}, 前場 \emph{maeba}, 相場 \emph{aiba}); (b) \emph{sound change}, where the correct morphemes are used but voicing or gemination is wrong (7 terms, e.g., 大手 \emph{\=ode}, 寄り付き \emph{yoridzuki}, 手仕舞い \emph{teshimai}, 日経平均 \emph{nichikei}); and (c) \emph{numeral euphony} (2 terms: 十分 \emph{j\=ubun}, 千円 \emph{sennen}). Type (a) changes several morae, whereas type (b) often changes a single phonetic feature. For each term we wrote two natural sentences in which context determines the financial reading, giving 60 sentences with a mean duration of 4.1\,s when synthesized.

\begin{table}[t]
\caption{Financial terms, expected readings and injected misreadings}
\label{tab:words}
\centering
\scriptsize
\setlength{\tabcolsep}{2.5pt}
\begin{tabular}{@{}lll|lll@{}}
\toprule
Word & Expected & Injected & Word & Expected & Injected \\
\midrule
市場 & しじょう & いちば & 決算 & けっさん & けつざん \\
大手 & おおて & おおで & 増益 & ぞうえき & ぞうやく \\
上値 & うわね & じょうね & 千円 & せんえん & せんねん \\
下値 & したね & しもね & 十分 & じゅっぷん & じゅうぶん \\
大引け & おおびけ & おおひけ & 人気 & にんき & ひとけ \\
相場 & そうば & あいば & 大勢 & たいせい & おおぜい \\
銘柄 & めいがら & めいへい & 一時 & いちじ & ひととき \\
手仕舞い & てじまい & てしまい & 後場 & ごば & あとば \\
円安 & えんやす & えんあん & 前場 & ぜんば & まえば \\
円高 & えんだか & えんこう & 空売り & からうり & くううり \\
為替 & かわせ & いがえ & 利食い & りぐい & りしょく \\
先物 & さきもの & せんもつ & 日経平均 & にっけいへいきん & にちけいへいきん \\
出来高 & できだか & できたか & 動向 & どうこう & どうむき \\
値幅 & ねはば & ちはば & 下落 & げらく & からく \\
押し目 & おしめ & おしもく & 寄り付き & よりつき & よりづき \\
\bottomrule
\end{tabular}
\end{table}

\subsection{Conditions and Ground Truth}
Each sentence was synthesized by each voice under three input conditions with a fixed seed: \textbf{K}, the target word replaced by its expected reading in hiragana; \textbf{W}, the target word replaced by the injected misreading in hiragana; and \textbf{N}, the original kanji text. This yields $60 \times 3 \times 3 = 540$ utterances. Because the front-end converts hiragana to phonemes deterministically, K and W utterances are requested to contain the correct and the wrong reading, respectively, which provides ground truth without manual labeling. The 360 K and W utterances form the \emph{controlled set}; the 180 N utterances measure behavior on natural input. In all conditions the verifier is given the original kanji sentence, as in production.

\subsection{Compared Methods}
We compare (i) the deployed Stage~1 gate; (ii) a direct phonetic transcription baseline, which decodes the hiragana CTC model greedily and flags the word if the expected reading does not occur in the output; and (iii) the proposed Stage~2 check with $\theta=0$. We report recall on W utterances, false-alarm rate on K utterances, precision, and the area under the ROC curve (AUC) of the margin $m(w)$, with Wilson 95\% confidence intervals and exact McNemar tests for paired comparisons.

\section{Results}

\subsection{Controlled Set}
Table~\ref{tab:main} summarizes the results. The deployed gate missed one in five injected misreadings and, more strikingly, rejected 44.4\% of utterances that contained the correct reading. Greedy phonetic transcription detected every misreading, but its output is noisy (e.g., ぼうひけ for おおびけ) and it rejected 48.3\% of correct utterances, which makes it unusable as a gate. The proposed check detected 91.7\% of misreadings with a 5.0\% false-alarm rate. Compared with Stage~1, it detected 35 misreadings that Stage~1 missed while missing 12 that Stage~1 caught ($p=0.001$), and it raised 3 false alarms that Stage~1 did not versus 74 in the opposite direction ($p<10^{-17}$).

\begin{table}[t]
\caption{Detection on the controlled set (3 voices, 180 W and 180 K utterances). Brackets: Wilson 95\% CI.}
\label{tab:main}
\centering
\scriptsize
\setlength{\tabcolsep}{3pt}
\begin{tabular}{@{}lccc@{}}
\toprule
Method & Recall (W) & False alarm (K) & Precision \\
\midrule
Stage 1: Whisper + kana & 78.9 [72.4, 84.2] & 44.4 [37.4, 51.7] & 64.0 \\
Greedy hiragana CTC & 100 [97.9, 100] & 48.3 [41.1, 55.6] & 67.4 \\
Proposed ($\theta=0$) & \textbf{91.7} [86.7, 94.9] & \textbf{5.0} [2.7, 9.2] & \textbf{94.8} \\
\bottomrule
\end{tabular}
\end{table}

\begin{figure}[t]
\centering
\includegraphics[width=\columnwidth]{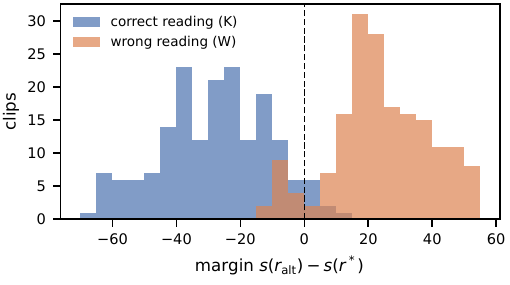}
\caption{Distribution of the margin $m(w)$ in \eqref{eq:margin} for utterances with the correct (K) and the injected wrong (W) reading. Dashed line: $\theta=0$.}
\label{fig:margin}
\end{figure}

\begin{figure}[t]
\centering
\includegraphics[width=0.92\columnwidth]{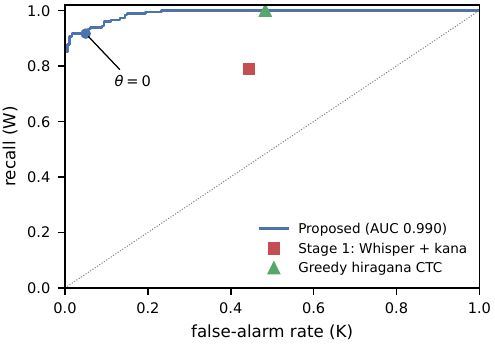}
\caption{ROC curve of the proposed margin on the controlled set, with the operating points of the two baselines.}
\label{fig:roc}
\end{figure}

Fig.~\ref{fig:margin} shows that the margin separates the two classes well (median $-26.9$ for K and $+23.1$ for W; AUC 0.990). The threshold offers a simple operating trade-off (Fig.~\ref{fig:roc} and Table~\ref{tab:theta}): $\theta=5$ gives 90.6\% recall at 1.7\% false alarms, and $\theta=-10$ gives 98.9\% recall at 15.0\% false alarms. Every operating point of the proposed check between these two dominates both baselines, which lie far to the right of the curve. Results were consistent across voices (Table~\ref{tab:voice}); AUC ranged from 0.983 to 0.998, with the best separation for the male voice.

\begin{table}[t]
\caption{Effect of the threshold $\theta$ (controlled set, 3 voices)}
\label{tab:theta}
\centering
\scriptsize
\begin{tabular}{@{}rcc@{}}
\toprule
$\theta$ & Recall (W) & False alarm (K) \\
\midrule
$-10$ & 98.9\% (178/180) & 15.0\% (27/180) \\
$-5$ & 93.9\% (169/180) & 8.3\% (15/180) \\
$0$ & 91.7\% (165/180) & 5.0\% (9/180) \\
$5$ & 90.6\% (163/180) & 1.7\% (3/180) \\
$10$ & 86.7\% (156/180) & 0.6\% (1/180) \\
\bottomrule
\end{tabular}
\end{table}

\begin{table}[t]
\caption{Per-voice recall / false-alarm rate (\%) and AUC of the margin}
\label{tab:voice}
\centering
\scriptsize
\begin{tabular}{@{}lccc@{}}
\toprule
Voice & Stage 1 & Proposed ($\theta=0$) & AUC \\
\midrule
F1 (female) & 81.7 / 50.0 & 93.3 / 6.7 & 0.988 \\
F2 (female) & 80.0 / 45.0 & 86.7 / 5.0 & 0.983 \\
M1 (male) & 75.0 / 38.3 & 95.0 / 3.3 & 0.998 \\
\bottomrule
\end{tabular}
\end{table}

\subsection{Why the Transcript-Based Gate Fails}
Of the 38 injected misreadings that passed Stage~1, Whisper wrote the target kanji word in 36. Table~\ref{tab:blind} lists typical cases for voice F1: the audio clearly contains the wrong reading, yet the transcript is the correctly written sentence, so its kana reading matches the input and the gate passes. The effect is strongest for the most common terms, where the language-model prior is strongest: Stage~1 detected misreadings of 相場 in only 1 of 6 utterances and of 市場, 後場 and 日経平均 in 2 of 6 each, whereas the proposed check detected 6, 5, 6 and 6 of 6 (Table~\ref{tab:word}).

\begin{table}[t]
\caption{Injected misreadings hidden by the recognizer (voice F1). The transcript restores the expected kanji, so Stage 1 passes.}
\label{tab:blind}
\centering
\scriptsize
\setlength{\tabcolsep}{2pt}
\begin{tabular}{@{}llr@{}}
\toprule
TTS input (injected reading) & Whisper transcript & $m(w)$ \\
\midrule
今週の東京\textbf{いちば}は、… & 今週の東京\textbf{市場}は… & $+19.3$ \\
金利の上昇が、\textbf{あいば}の重しに… & 金利の上昇が\textbf{相場}の重しに… & $+28.7$ \\
取引開始から\textbf{じゅうぶん}後に、… & 取引開始から\textbf{十分}後に、… & $+32.1$ \\
\textbf{あとば}は日銀の発表待ちで、… & \textbf{後場}は日銀の発表待ちで、… & $+24.7$ \\
\textbf{にちけいへいきん}は、三日ぶりに… & \textbf{日経平均}は3日ぶりに… & $+16.4$ \\
原油価格の\textbf{からく}が、… & 原油価格の\textbf{下落}が… & $+22.9$ \\
\bottomrule
\end{tabular}
\end{table}

\begin{table}[t]
\caption{Detections per term (out of 6 utterances: 3 voices $\times$ 2 sentences). W: misreadings detected; K: false alarms.}
\label{tab:word}
\centering
\scriptsize
\setlength{\tabcolsep}{2.2pt}
\begin{tabular}{@{}lcccc|lcccc@{}}
\toprule
& \multicolumn{2}{c}{Stage 1} & \multicolumn{2}{c|}{Proposed} & & \multicolumn{2}{c}{Stage 1} & \multicolumn{2}{c}{Proposed} \\
Term & W & K & W & K & Term & W & K & W & K \\
\midrule
市場 & 2 & 3 & 5 & 0 & 決算 & 5 & 1 & 6 & 0 \\
大手 & 3 & 1 & 5 & 0 & 増益 & 6 & 3 & 6 & 0 \\
上値 & 6 & 3 & 6 & 0 & 千円 & 6 & 1 & 6 & 0 \\
下値 & 6 & 1 & 6 & 0 & 十分 & 3 & 1 & 6 & 1 \\
大引け & 3 & 2 & 6 & 3 & 人気 & 3 & 4 & 5 & 0 \\
相場 & 1 & 0 & 6 & 0 & 大勢 & 5 & 6 & 4 & 0 \\
銘柄 & 6 & 4 & 6 & 0 & 一時 & 6 & 0 & 6 & 0 \\
手仕舞い & 6 & 6 & 6 & 3 & 後場 & 2 & 4 & 6 & 0 \\
円安 & 5 & 2 & 4 & 0 & 前場 & 5 & 4 & 6 & 0 \\
円高 & 6 & 2 & 6 & 1 & 空売り & 5 & 3 & 4 & 0 \\
為替 & 6 & 3 & 6 & 0 & 利食い & 6 & 3 & 6 & 0 \\
先物 & 6 & 5 & 6 & 0 & 日経平均 & 2 & 0 & 6 & 1 \\
出来高 & 6 & 3 & 6 & 0 & 動向 & 6 & 2 & 6 & 0 \\
値幅 & 6 & 6 & 6 & 0 & 下落 & 4 & 1 & 6 & 0 \\
押し目 & 6 & 2 & 6 & 0 & 寄り付き & 4 & 4 & 0 & 0 \\
\bottomrule
\end{tabular}
\end{table}

Stage~1 false alarms on correct audio were caused by recognition errors on jargon and homophones that change the kana string: 上値 transcribed as お金, 手仕舞い as 手締い, 大引け as 大引き, 円高 as 縁どこ. In production, each such false alarm costs a regeneration and, when all candidates fail, leaves a correct sentence marked as suspicious.

\subsection{Failure Cases of the Proposed Check}
The proposed check missed all six misreadings of 寄り付き (\emph{yoritsuki} vs.\ \emph{yoridzuki}). The two readings differ only in the voicing of one mora, and the CTC model's outputs for this region were unstable (e.g., \emph{yarisuki}, \emph{yoritsuki}) in both conditions. Excluding this term, recall was 94.8\% (165/174) at the same false-alarm count. The remaining misses (円安, 空売り, 大勢, 市場, 大手, 人気; one or two utterances each) had small negative margins in utterances where the recognizer's output was degraded overall. The nine false alarms concentrated on 大引け and 手仕舞い, terms whose correct reading contains a voiced mora that the recognizer frequently heard as unvoiced. These cases suggest that minimal voicing contrasts call for a stronger acoustic model or a word-specific threshold.

Table~\ref{tab:type} confirms this pattern by misreading type. For reading-choice errors, which change several morae, the proposed check detected 93.7\% with a single false alarm in 126 correct utterances, while Stage~1 produced 61 false alarms. Sound-change errors were harder for both methods: the proposed check detected 83.3\% (35/42), six of the seven misses being 寄り付き, and 7 of its 9 false alarms fell into this type. Numeral-euphony errors were detected in all 12 utterances.

\begin{table}[t]
\caption{Recall / false alarms by misreading type (controlled set)}
\label{tab:type}
\centering
\scriptsize
\setlength{\tabcolsep}{3pt}
\begin{tabular}{@{}lccc@{}}
\toprule
Type (terms, utterances per class) & Stage 1 & Proposed \\
\midrule
(a) Reading choice (21, 126) & 104/126, 61 FA & 118/126, 1 FA \\
(b) Sound change (7, 42) & 29/42, 17 FA & 35/42, 7 FA \\
(c) Numeral euphony (2, 12) & 9/12, 2 FA & 12/12, 1 FA \\
\bottomrule
\end{tabular}
\end{table}

\subsection{Combining the Two Stages}
In production the two stages are applied in sequence, so a sentence is rejected if either fails. With the deployed exact-match Stage~1, the combination detects 177/180 misreadings but inherits Stage~1's false alarms (83/180 on K). Stage~1 is still needed for omissions, insertions and garbled speech, which Stage~2 does not target, but it need not require an exact kana match. Relaxing Stage~1 to reject only sentences whose kana similarity to the input (difflib ratio) falls below 0.8 leaves just 2 Stage~1 rejections among the 180 correct K utterances; combined with Stage~2 this yields 165/180 recall at 11/180 false alarms, and 12/180 regeneration requests on natural input. We therefore recommend a lenient Stage~1 for gross errors and Stage~2 for readings. Our test set contains no omissions, so the appropriate Stage~1 threshold for catching them remains to be validated.

\subsection{Computational Cost}
Stage~2 requires one forward pass of the 300M-parameter CTC model per sentence and one CTC loss per candidate reading. On one NVIDIA H100 NVL GPU, scoring took 17.6\,ms per sentence on average (median 14.9\,ms) for utterances averaging 4.1\,s, a real-time factor of 0.004, which is negligible compared with synthesis and Whisper transcription.

\subsection{Natural Input}\label{sec:natural}
With the original kanji input, the front-end produced the expected reading for 29 of 30 terms; it read 大勢 as \emph{\=oze} (a crowd) in both of its sentences, so all six N utterances of 大勢 contained a real misreading originating in the front-end. Stage~1 flagged 4 of the 6: for M1 both utterances passed because Whisper transcribed the audio as 大勢, which pyopenjtalk again read as \emph{\=oze}, matching the (equally wrong) expected reading derived from the input. The proposed check, whose expected reading comes from the domain lexicon rather than the front-end, flagged 5 of 6. This illustrates a second blind spot: a verifier that derives its expectation from the same front-end as the synthesizer cannot detect front-end errors.

On the remaining 174 natural utterances, Stage~1 requested regeneration for 66 (37.9\%) and the proposed check for 6 (3.4\%); over all 180 natural utterances the rates were 38.9\% and 6.1\%. Stage~1's rate varied strongly with the voice (F1 50.0\%, F2 41.7\%, M1 25.0\%), reflecting how well Whisper recognized each voice rather than how many misreadings it produced; the proposed check's rate was 10.0\%, 5.0\% and 3.3\%, respectively. The six natural utterances flagged only by the proposed check (大引け $\times 2$, 手仕舞い $\times 3$, 一時 $\times 1$) were verified by a native listener, and all six contained the correct reading. They are therefore false alarms, and they involve the same voiced morae (\emph{bi} in \emph{\=obike}, \emph{ji} in \emph{tejimai}) that caused false alarms in the controlled set. In total, 5 of the proposed check's 11 alarms on natural input were genuine front-end misreadings and 6 were false alarms, a false-alarm rate of 3.4\% (6/174) on utterances without front-end errors, consistent with the 5.0\% measured on the controlled set. Stage~1 raised 66 alarms on the same 174 utterances; we did not verify all of them by listening, but the front-end readings of these utterances were correct, 4 of them are among the six utterances verified as correct above, and the remaining 62 had negative margins, so most of them are likely needless regenerations.

\begin{figure}[t]
\centering
\includegraphics[width=\columnwidth]{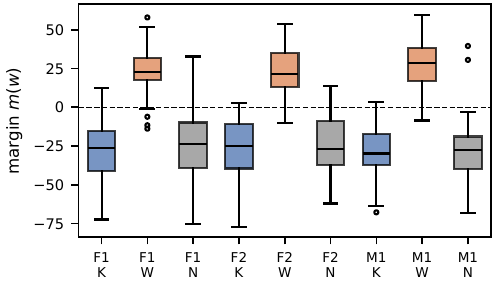}
\caption{Margin $m(w)$ per voice and condition (K: correct reading injected, W: wrong reading injected, N: natural kanji input). Natural utterances behave like K except for the front-end misreadings of 大勢, which appear as positive outliers.}
\label{fig:voice}
\end{figure}

\section{Discussion and Limitations}
\textbf{Practical guidelines.} From these results we derive three recommendations for production pipelines. (1) Do not use exact kana match of an ASR transcript as the only correctness gate for Japanese: it both hides misreadings of common terms and rejects many correct sentences. Use it leniently, for gross errors only. (2) Verify readings of known ambiguous terms with likelihood comparison of explicit candidates, with the expected reading taken from a domain lexicon rather than from the synthesizer's front-end. (3) Choose $\theta$ according to the cost of errors: for published content where a misreading is costly, a negative threshold (e.g., $\theta=-5$, 93.9\% recall at 8.3\% false alarms) is affordable because a false alarm only costs one regeneration of a short sentence. Fig.~\ref{fig:voice} also shows that natural utterances score like the K condition, so the controlled false-alarm rates are representative of production.

\textbf{Limitations.}
The controlled set relies on injected misreadings, which are spoken fluently by the TTS model and may be easier to detect than natural acoustic deviations that are partially articulated. The ground truth assumes that the acoustic model followed the requested phonemes; the K-condition false alarms show that this is mostly but not always true, so a small part of the reported errors may be label noise. The proposed check requires a lexicon of terms with their expected and confusable readings. In our use case this lexicon grows naturally from the substitution table and from domain glossaries, but it does not cover unseen words; Stage~1 remains necessary for omissions and insertions, and the two stages are complementary. Finally, the study uses one TTS architecture, one CTC model trained on read speech, and 60 sentences; pitch-accent errors, which are also noticeable to native listeners, are not addressed.

\section{Conclusion}
We showed that ASR-based verification of synthesized Japanese, even at the kana level, is unreliable for kanji reading errors: the recognizer's language model hides misreadings of common financial terms, recognition errors on jargon cause frequent false alarms, and expectations derived from the synthesizer's own front-end cannot reveal front-end errors. A reading-level check that compares CTC likelihoods of the expected and confusable readings detected 91.7\% of injected misreadings at a 5.0\% false-alarm rate across three voices, caught a real front-end misreading, and reduced regeneration requests on natural input from 38.9\% to 6.1\%. Future work includes automatic generation of confusable readings, word-specific thresholds for voicing contrasts, pitch-accent verification, and extension to Mandarin polyphonic characters.


\begin{thebibliography}{99}
\bibitem{valle}
C. Wang \emph{et al.}, ``Neural codec language models are zero-shot text to speech synthesizers,'' arXiv:2301.02111, 2023.

\bibitem{speartts}
E. Kharitonov \emph{et al.}, ``Speak, read and prompt: High-fidelity text-to-speech with minimal supervision,'' \emph{Trans. Assoc. Comput. Linguist.}, vol.~11, pp.~1703--1718, 2023.

\bibitem{lora}
E. J. Hu \emph{et al.}, ``LoRA: Low-rank adaptation of large language models,'' in \emph{Proc. ICLR}, 2022.

\bibitem{openjtalk}
Open JTalk: HMM-based Japanese text-to-speech system. [Online]. Available: \url{https://open-jtalk.sourceforge.net/}

\bibitem{pyopenjtalk}
R. Yamamoto, ``pyopenjtalk: A Python wrapper for OpenJTalk,'' GitHub repository. [Online]. Available: \url{https://github.com/r9y9/pyopenjtalk}

\bibitem{g2pw}
Y.-C. Chen, Y.-C. Chang, Y.-C. Chang, and Y.-R. Yeh, ``g2pW: A conditional weighted softmax BERT for polyphone disambiguation in Mandarin,'' in \emph{Proc. Interspeech}, 2022, pp.~1926--1930.

\bibitem{vits}
J. Kim, J. Kong, and J. Son, ``Conditional variational autoencoder with adversarial learning for end-to-end text-to-speech,'' in \emph{Proc. ICML}, PMLR vol.~139, 2021, pp.~5530--5540.

\bibitem{whisper}
A. Radford, J. W. Kim, T. Xu, G. Brockman, C. McLeavey, and I. Sutskever, ``Robust speech recognition via large-scale weak supervision,'' in \emph{Proc. ICML}, PMLR vol.~202, 2023, pp.~28492--28518.

\bibitem{whisperx}
M. Bain, J. Huh, T. Han, and A. Zisserman, ``WhisperX: Time-accurate speech transcription of long-form audio,'' in \emph{Proc. Interspeech}, 2023, pp.~4489--4493.

\bibitem{gop}
S. M. Witt and S. J. Young, ``Phone-level pronunciation scoring and assessment for interactive language learning,'' \emph{Speech Communication}, vol.~30, no.~2, pp.~95--108, 2000.

\bibitem{ctc}
A. Graves, S. Fern\'andez, F. Gomez, and J. Schmidhuber, ``Connectionist temporal classification: Labelling unsegmented sequence data with recurrent neural networks,'' in \emph{Proc. ICML}, 2006, pp.~369--376.

\bibitem{xlsr}
A. Conneau, A. Baevski, R. Collobert, A. Mohamed, and M. Auli, ``Unsupervised cross-lingual representation learning for speech recognition,'' in \emph{Proc. Interspeech}, 2021, pp.~2426--2430.

\bibitem{commonvoice}
R. Ardila \emph{et al.}, ``Common Voice: A massively-multilingual speech corpus,'' in \emph{Proc. LREC}, 2020, pp.~4218--4222.

\bibitem{w2v2hira}
vumichien, ``wav2vec2-large-xlsr-japanese-hiragana,'' Hugging Face model, 2022, doi:10.57967/hf/0344. [Online]. Available: \url{https://huggingface.co/vumichien/wav2vec2-large-xlsr-japanese-hiragana}
\end{thebibliography}
\end{document}